\documentclass[a4paper,11pt]{article}
\usepackage{graphicx,times}
\usepackage{natbib}
\bibpunct{(}{)}{;}{a}{}{,}

\begin{document}

\title{Trailing (L5) Neptune Trojans:\\ 2004 KV18 and 2008 LC18}

\author{Pu Guan, Li-Yong Zhou\footnote{Corresponding author, {\it
   zhouly@nju.edu.cn}}, Jian Li \\
Astronomy Department, Nanjing University, \\ Key Laboratory
of Modern Astronomy and Astrophysics in MOE,\\
 Nanjing 210093, China }

\maketitle

\begin{abstract}
The population of Neptune Trojans is believed to be bigger than that
of Jupiter Trojans and that of asteroids in the main belt, although
only eight members of this far distant asteroid swarm have been
observed up to now. Six leading Neptune Trojans around the Lagrange
point $L_4$ discovered earlier have been studied in detail, but two
trailing ones found recently around the $L_5$ point, 2004 KV18 and
2008 LC18, have not been investigated yet. In this paper, we report
our investigations on the dynamical behaviors of these two new
Neptune Trojans. Our calculations show that the asteroid 2004 KV18
is a temporary Neptune Trojan. Most probably, it was captured into
the trailing Trojan cloud no earlier than $2.03\times10^5$\,yr ago,
and it will not keep this identity no later than
$1.65\times10^5$\,yr in future. Based on the statistics on our
orbital simulations, we argue that this object is more like a
scattered Kuiper belt object. On the contrary, the orbit of asteroid
2008 LC18 is much more stable. Among the clone orbits spread within
the orbital uncertainties, a considerable portion of clones may
survive on the $L_5$ tadpole orbits for 4\,Gyr. The strong
dependence of the stability on the semimajor axis and resonant angle
suggests that further observations are badly needed to confine the
orbit in the stable region. We also discuss the implications of the
existence and dynamics of these two trailing Trojans on the Solar
system history.
\end{abstract}

\section{Introduction}           
\label{sect:intro}

The Trojans are celestial bodies moving on the same orbit as a
planet, but around $60^{\circ}$ ahead or $60^{\circ}$ behind the
planet close to the triangular Lagrange points $L_4$ (leading) or
$L_5$ (trailing). By the original definition, only those asteroids
on the so-called tadpole orbits are ``real'' trojans \citep{mur99}.
Jupiter is the first planet known to host thousands of this kind of
asteroids after the discovering of (588) Achilles in 1906. Several
Trojan asteroids around Mars were discovered quite lately in 1990s
\citep{bow90}. Even another ten years later, the first Neptune
Trojan 2001 QR322 was found \citep{chi03} to orbit around the $L_4$
Lagrange point. And in the August of 2011, the first Earth Trojan
was confirmed \citep{mai11} and its dynamics was studied
\citep{con11,dlz12} very recently.

The trojan asteroids are of special interest not only because the
dynamics of them is complicated, but also because their origin and
evolution may bear important clues to the early history of our Solar
system. Many studies, e.g. \cite{nes02a, mar03, rob06, rob09, dvo07,
dvo10, zly1, zly2}, have devoted to the dynamics of Trojan asteroids
around different planets. In recent years, the origin of Trojans and
the formation of the Trojans cloud began to attract more and more
attentions \citep{mor05, nes09, lyk09, lyk10a} since the very
well-known ``Nice Model'' \cite[for a review, see for
example][]{cri09} about the early history of the Solar system
regards the existence and property of Jupiter trojans as one of the
critical evidences of the theory \citep{mor05}.

Before the discovery of asteroid 2008 LC18 by \cite{she10a}, six
Neptune Trojans (NT hereafter for short) have been
observed\footnote{See the website of IAU: Minor Planet Center with
URL http://www.minorplanetcenter.net/iau/lists/NeptuneTrojans.html
}. But they are all around the leading Lagrange point $L_4$, about
$60^\circ$ ahead of Neptune. This is partly due to the fact that the
trailing Lagrange point ($L_5$) is nowadays in the Galaxy center
direction. The background stars add difficulty to the discovery of
asteroids in this ``shinning'' region. As the first NT orbiting
around the $L_5$ point, 2008 LC18 is of particular interest, not
only because it is the first member of a possible asteroid swarm in
which it resides, but also because a tricky way has been used to
block out the strong background light from the Galaxy center.
Following this success, another $L_5$ NT (2004 KV18) was reported in
July of 2011, adding the number of $L_5$ NTs to two. As for orbital
dynamics researching, these two findings make the first step of
confirming the dynamical symmetry between the $L_4$ and $L_5$ points
\citep{nes02a,mar03,zly1}, and the high inclinations of their orbits
(see Table~\ref{orbele}) raise further the fraction of NTs on
highly-inclined orbits (so far 3 out of total 8 NTs have inclination
larger than 25 degrees). Both of these two points bring specific
indications to the capturing process of NTs and the evolution of the
planetary system in the early stage \citep{nes09,lyk09,lyk10a}.
Meanwhile, the New
Horizons\footnote{http://www.nasa.gov/mission\_pages/newhorizons/main/index.html}
probe will travel through the sky region around Neptune's $L_5$
point in a few years, thus the study of this region is even more
important than the one around $L_4$.

People meet difficulties in explaining the estimated 4:1 high
inclination excess among the population of NTs \citep{she06}.
Investigations on their dynamics show that the inclination of NTs is
not likely excited {\it in situ} under the current planetary
configuration. The only acceptable explanation seems to be that the
NTs are captured rather than formed {\it in situ}, and the capture
progress pumped up the Trojans' orbits, resulting in both high
inclination and high eccentricity \citep{nes09,lyk10b}. On the other
hand, the NT orbit with eccentricity larger than 0.1 seems to be
unstable, thus the NTs excited in the early days of the Solar system
should have been expelled from the Trojan cloud \citep{zly1,zly2}.
This is the critical puzzle in the ``capture origin'' scenario.

Except for the newly found asteroid 2004 KV18, all other NTs have
eccentricities below 0.1. The highly eccentric orbit ($e = 0.184$)
makes it so peculiar. Can it be the ``smoking gun'' to support the
capture origin of Neptune Trojan cloud? Or, it is not a remnant from
the original dynamically excited Trojan cloud, but just a passer-by
on its journey from the trans-Neptune region to the inner
interplanetary space. We take great interest to explore its
dynamical properties to find some clues.

In this paper, we study in detail the dynamics of these two $L_5$
Neptune Trojans. The paper is organized as follow. We give our model
and method of numerical simulations in Section 2. The simulation
results are summarized and discussed in Section 3 for 2004 KV18 and
in Section 4 for 2008 LC18. Finally, we make the conclusions in
Section 5.

\section{Model and Method}
\label{sect:M&M}

The asteroids 2008 LC18 and 2004 KV18 have been observed at
opposition for 2 and 3 times respectively, and their orbits have
been determined. Their orbital elements, taken from the AstDyS ({\it
Asteroids - Dynamic Site})
website\footnote{http://hamilton.dm.unipi.it}, are listed in
Table~\ref{orbele}. The uncertainties arising from the observation
and orbital determination processes, are listed in the Table as
well. In our previous papers \citep{zly1,zly2}, we have constructed
dynamical maps and resonant maps on the $(a,i)$ and $(a,e)$ planes
to show the locations of important resonances that control the
dynamics of NTs. If we find the positions of these two objects on
the corresponding maps, we may get immediately the conjecture about
their dynamical behaviors. For this sake, we transfer the orbital
elements of these two objects to the right epoch JD=2449200.5 at
which the maps were composed. They are given in Table~\ref{orbele}
too.

\begin{table}
\caption{Orbital elements of asteroids 2008 LC18 and 2004 KV18,
given at epoch JD=2455800.5 (2011-Aug-27) with respect to the mean
ecliptic and equinox at J2000. The semimajor axes are given in AU,
while the angular elements including inclination $i$, perihelion
argument $\omega$, ascending node $\Omega$ and mean anomaly $M$ are
in degrees.The elements and $1\sigma$ variations are taken from the
AstDyS (see text). For the sake of comparing with our previous
results, the orbital elements at epoch JD=2449200.5 (see text for
explanation) are listed in columns indicated by ``Val4Com''. }
 \begin{center}
 \begin{tabular}{|c|lll|lll|}
 \hline
  & \multicolumn{3}{c}{2008 LC18}  \vline & \multicolumn{3}{c}{2004 KV18} \vline \\
 \cline{2-7}
  \raisebox{1.6ex}[0pt]{Elements} & Value & $1\sigma$ & Val4Com & value & $1\sigma$ & Val4Com \\
 \hline
  $a$ & 29.9369 & 0.02588 & 30.1010 & 30.1260 & 0.01088 & 30.3927\\
  $e$ & 0.083795 & 0.002654 & 0.080360 & 0.183846 & 0.000797 & 0.190021 \\
  $i$ & 27.5689  & 0.003824 & 27.5144 & 13.6092 & 0.001336 & 13.5684 \\
  $\Omega$ & 88.521  & 0.0007854 & 88.549 & 235.6273 & 0.0004537 & 235.6852\\
  $\omega$ & 5.1349 & 10.85 & 8.9773 & 294.5615 & 0.1789 & 295.7312 \\
  $M$ &   173.909 & 12.83 & 130.057 & 58.5939 & 0.09 &18.1013 \\
 \hline
 \end{tabular}
 \end{center}
 \label{orbele}
\end{table}

With the orbital elements of these objects, we perform numerical
simulations to investigate their orbital evolutions and orbital
stabilities. We adopt the Outer Solar System model, namely the
gravitational system consisting of the Sun and four jovian planets
from Jupiter to Neptune. The planets are in their current orbits and
the NTs are assumed to be zero-mass particles. Taking into account
the uncertainties in the orbital elements, to make our
investigations convincing, it is necessary to consider some clone
orbits around the nominal orbits of these objects. Using the
covariance matrix given by the AstDyS website, we generate for each
object a cloud of 1000 clone orbits in the 6-dimensional orbital
elements space centered on the nominal orbit. The distribution of
these elements, for 2008 LC18 as an example, can be seen in
Fig.~\ref{lcini}. The orbits are numerically simulated using the
hybrid algorithm from the {\it Mercury6} package \citep{cham99}.

\section{2004 KV18}
\label{sect:kv18}

For the asteroid 2004 KV18, 1000 clone orbits are integrated up to
10~Myr in both forward (to the future) and backward (to the past)
directions. The integration time span (10~Myr) is chosen after some
test computations, and it is long enough to show the behavior of the
asteroid as being an $L_5$ Neptune Trojan. In fact, according to the
results in our previous papers \citep{zly1,zly2}, the orbit of an NT
can be stable only when its eccentricity is smaller than 0.12. The
location of the nominal orbit of 2004 KV18 on the dynamical map at
epoch JD=2449200.5 \citep[figure~3 in][]{zly2}, i.e.
$(a,e)=(30.3927, 0.190021)$ as listed in Table~\ref{orbele}, reveals
that this orbit locates far away from the stable region. Note that
we did not show the dynamical map at the section of the exact
inclination $13.5684^\circ$, but we had the maps for $i=10^\circ,
20^\circ$ in the paper and for $i=15^\circ$ not presented in the
paper, and the continuity helps us draw the above conclusion.

\subsection{Lifespan as an $L_5$ Neptune Trojan}

We have removed those clones whose semimajor axes exceed 100\,AU in
our simulations, because they have been practically ejected from the
Solar system. By the end of the integration (10\,Myr), 569 clones
out of the total 1000 in the forward integrations, and 558 in the
backward integration, survive (stay in the Solar system). The
leftovers, however seem all on the very chaotic orbits and we
believe they will escape the Solar system some time. Since our
interests in this paper are mainly in the fate of the asteroid as a
Neptune Trojan, we track the variable $\sigma$ defined as:
 \begin{displaymath}
 \sigma = \lambda - \lambda_8
 \end{displaymath}
\noindent where $\lambda$ and $\lambda_8$ are the mean longitudes of
the clone and Neptune, respectively. So $\sigma$ is the critical
argument of the 1:1 mean motion resonance (MMR) between the clone
and Neptune. For an $L_5$ trojan, $\sigma$ librates around
$-60^\circ$ (or equivalently $300^\circ$). When the librating
amplitude\footnote{In this paper, ``amplitude''
refers to the full range of $\sigma$ variation, i.e. the difference
between the maximal and minimal $\sigma$ values.} is not so large that $-180^\circ < \sigma <0^\circ$, the
trojan is on a ``tadpole orbit''. If the amplitude gets larger
resulting in $\sigma < -180^\circ$, the asteroid may still stay in
the 1:1 MMR but turn to a ``horseshoe orbit''. It is not a
``Trojan'' anymore according to the original definition
\citep{mur99}. And if $\sigma$ circulates, the asteroid leaves the
1:1 MMR. We define the time $t_1$, when for the first time $\sigma$
goes beyond $-180^{\circ}$ in our simulations, as the moment of a
clone escaping from the Trojan cloud. When a clone begins to
circulate for the first time at moment $t_2$, it is regarded as
having escaped from the 1:1 MMR.

\begin{figure}
   \centering
   \includegraphics[width=\textwidth, height=6.5cm,angle=0]{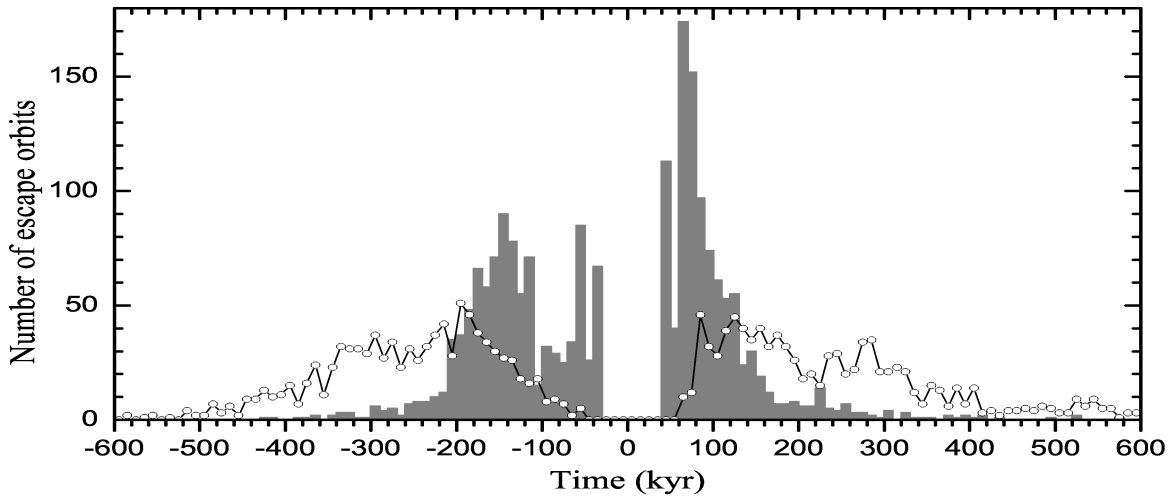}
   \caption{Number of escaping orbit per 10\,kyr. The dark grey histogram
   in the background show the distribution of moments $t_1$ when the clones
   leave the $L_5$ Trojan region (see text), while the curves with open
   circles represent the distribution of $t_2$ when the clone escape from
   the 1:1 MMR. }
   \label{kvesct}
   \end{figure}

We summarize in Fig.~\ref{kvesct} the distributions of $t_1$ and
$t_2$ for the 1000 clones in both time directions. Judged by $t_1$,
none of the clones keeps the $L_5$ NT identity in 10\,Myr in both
time directions. Moreover, all of the clones leave the 1:1 MMR in
our simulations (although some of them were recaptured into 1:1
resonance again by the end). This orbital instability is consistent
with the conclusion drawn from locating the nominal orbit on the
dynamical maps in our previous papers \citep{zly1,zly2}.

By checking carefully the data, we find that the first escaping from
the Trojan orbit ($\sigma < -180^\circ$) in the forward integrations
happens at $\sim 46$\,kyr, the escaping orbits pile up quickly
making the first peak at $\sim 48$\,kyr. And soon after, the highest
peak appears around $\sim 73$\,kyr. Till $\sim 165$\,kyr, 90\%
clones have left the $L_5$ region. For the backward integrations,
the distribution of $t_1$ is more flat than that in the forward
direction. The first escaping happens $\sim -34$\,kyr, and the first
peak appears here. Very closely, another peak emerges at $\sim
-58$\,kyr. About 30\% of escapes are hosted by these two peaks. And
a wide peak centered at $\sim -155$\,kyr contains nearly all the
rest clones. Till $-203$\,kyr, 90\% clones have escaped from the
$L_5$ tadpole orbits. Based on the above statistics, we may conclude
confidently that the asteroid 2004 KV18 has been an $L_5$ Neptune
Trojan for at least 34\,kyr and it will keep this identity for at
least 46\,kyr in future. But most probably, it's neither a
primordial nor a permanent member of the $L_5$ NT cloud\footnote{By
``primordial'' we mean that the asteroid has been in the Trojan
region since very early, before or just after the planets attained
their current orbits, no matter the asteroid was captured by or
grown up with Neptune. By ``permanent'', we mean that the asteroid
will stay in the Trojan region until 4\,Gyr later.}. With a
probability of 90\%, it became an $L_5$ NT no earlier than 203\,kyr
ago and it will leave the $L_5$ region in less than 165\,kyr.

Soon after leaving the tadpole orbits around the $L_5$ point, the
clones escape the 1:1 MMR in both time directions, as shown by the
$t_2$ distribution in Fig.~\ref{kvesct}. In fact, after leaving the
$L_5$ tadpole orbit and before its escaping from the MMR, a clone
orbit may experience the horseshoe orbit, enter the tadpole orbit
around the $L_4$ point, or even become a retrograde satellite of
Neptune, but all these experiences only last for very short time. We
will show some detail below.

\subsection{Orbital evolution}

\begin{figure}
  \centering
   \includegraphics[width=\textwidth, angle=0]{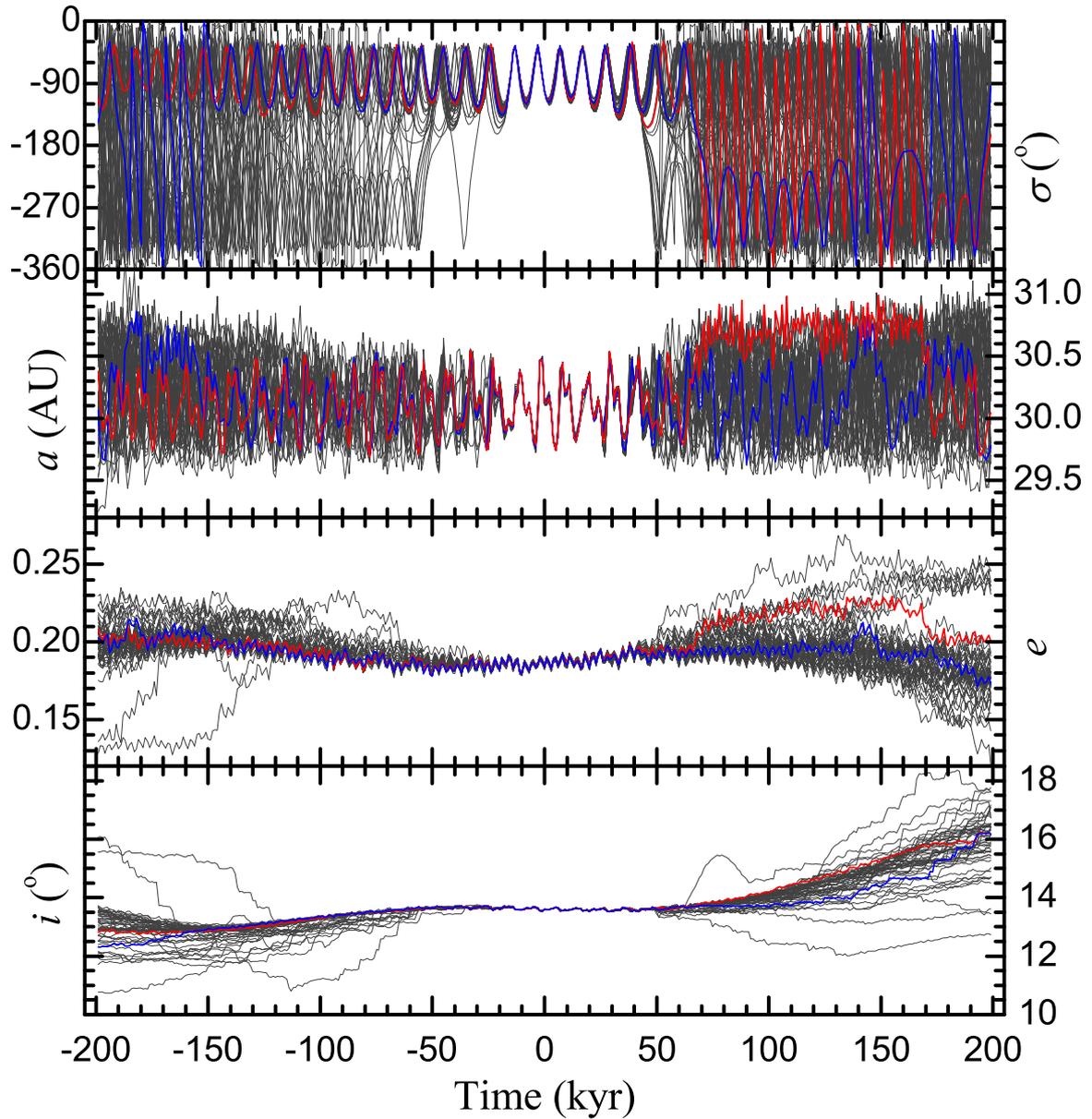}
   \caption{From top to down, we show the temporal evolution of the resonant angle ($\sigma$),
   semimajor axis ($a$), eccentricity ($e$) and inclination ($i$)
   of 50 clones. The nominal orbit is represented by the thick and red
   curves. And another arbitrarily selected orbit is highlighted and plotted in blue (see text).
   }
  \label{kvorb}
\end{figure}

To show the temporal evolution of the clone orbits, we plot in
Fig.~\ref{kvorb} the resonant angle ($\sigma$), semimajor axis
($a$), eccentricity ($e$), and inclination ($i$) of 50 clone orbits.
These clones are selected arbitrarily (except the nominal orbit)
from our 1000 samples. Clearly the orbits are chaotic, as the
initially close-to-each-other orbits at time $t=0$ become separated
quite soon.

In the top panel, we see that the resonant angle $\sigma$ librates
for several periods in both forward and backward directions before
it's libration amplitude increases beyond
$180^\circ$ where the clone turns to the horseshoe orbit. The
libration period $T$ of a Trojan can be estimated through \citep[see
for example][]{mur99}
 \begin{displaymath}
 T= 2\pi / \sqrt{\frac{27}{4}{\mu}}
 \end{displaymath}
\noindent where $\mu$ is the mass of Neptune with respect to the
Solar mass. This equation is only valid for tadpole orbits in the
very vicinity of $L_{4,5}$ points, and it leads to a period of
8.9\,kyr for Neptune Trojan. The 2004 KV18 is quite far away from
the $L_5$ point with a $\sigma$ amplitude $\sim 100^\circ$
(see Fig.~\ref{kvorb}), and the libration period is $\sim
10.7$\,kyr. A clone always leaves the $L_5$ region just after
$\sigma$ reaches the minimal value, and this explains the abruptly
appearing peaks in Fig.~\ref{kvesct} at $t\sim -34$\,kyr and $t\sim
48$\,kyr. Between these two peaks, $\sigma$ finishes 7 whole
librating periods. The libration amplitude must be tuned by some
periodic effects (e.g. secular resonances), that's the reason some
periodic features can be seen in the distribution of $t_1$ and $t_2$
in Fig.~\ref{kvesct}.

After leaving the $L_5$ region in the future (or before being
captured in the past), the clone may have
different experiences before it's final escaping from the 1:1 MMR.
It may librate with an amplitude larger than $180^\circ$, moving on
a horseshoe orbit; it may shift to the tadpole orbit around the
$L_4$ point and become a leading Trojan; and it may behave like a
retrograde satellite around Neptune with $\sigma$ librating around
$0^\circ$ with a small amplitude. The nominal orbit and another
highlighted orbit in Fig.~\ref{kvorb} show clearly some of these
possibilities. In a word, the asteroid 2004 KV18 is an $L_5$ Neptune
Trojan right now, but probably it was and will be in the $L_4$
Trojan cloud, and it may change its identity several times in the
time duration of being in the 1:1 MMR.

When a clone is on the tadpole orbit (either around the $L_5$ or
$L_4$ point), its orbital evolution is more or less regular in this
regime. For example, in the time range from $-200$\,kyr to 70\,kyr,
the nominal orbit is an $L_5$ Trojan orbit as indicated by
$\sigma$'s behavior, and we find its semimajor axis, eccentricity
and inclination all behave regularly, as shown in
Fig.~\ref{kvorb}. For another highlighted orbit (blue curves in
Fig.~\ref{kvorb}), the tadpole stage is from $-150$\,kyr to 140\,kyr
($L_5$ first and $L_4$ later, shifting at $\sim 70$\,kyr), again the
regular evolutions in $a, e$ and $i$ can be clearly recognized.

But the regular motion must be a transient phenomenon, because the
orbit locates on the separatrix of the 1:1 MMR. This region is
characterized by strong chaos induced by overlaps of the secondary
resonances \citep[see for example][]{mit95}. Moreover, comparing the
orbital elements with the resonance map in our previous papers
(figure 11 in \cite{zly1} and figure 5 in \cite{zly2}), we see that
the 2004 KV18 may be strongly influenced by a combined resonance
characterized by $2f_\sigma - f_{2:1} + g_6 =0$, where $f_\sigma$
and $f_{2:1}$ are the frequencies of the resonant angle $\sigma$ and
the quasi 2:1 MMR between Neptune and Uranus, and
$g_6$ is the apsidal precession rate of Saturn.

As soon as the clone leaves 1:1 MMR, the orbital evolution will be
very chaotic. This chaos arises from the overlap of one-order MMRs
in the close vicinity of a planet's orbit
\citep{wis80,dun89}. Meanwhile, because Neptune and Uranus are very
close to the 2:1 MMR, some clones will enter the 2:1 MMR with Uranus
immediately after it escapes the 1:1 MMR with Neptune. In fact the
nominal orbit is in the 2:1 MMR with Uranus from 70\,kyr to 170\,kyr
when its $a$ librates around 30.7\,AU with a small amplitude, as
shown in the second panel of Fig.~\ref{kvorb}.

Due to the strong chaos, it is hard to predict precisely the
long-term fate of the clones after they escape from the 1:1 MMR,
or inversely in time, to trace back their origins before being
capturing into this 1:1 MMR. However, we can still make some
informative statistics on their evolutions as follow.

\subsection{Past and future}

\begin{figure}
   \centering
   \includegraphics[width=\textwidth, height=6.5cm, angle=0]{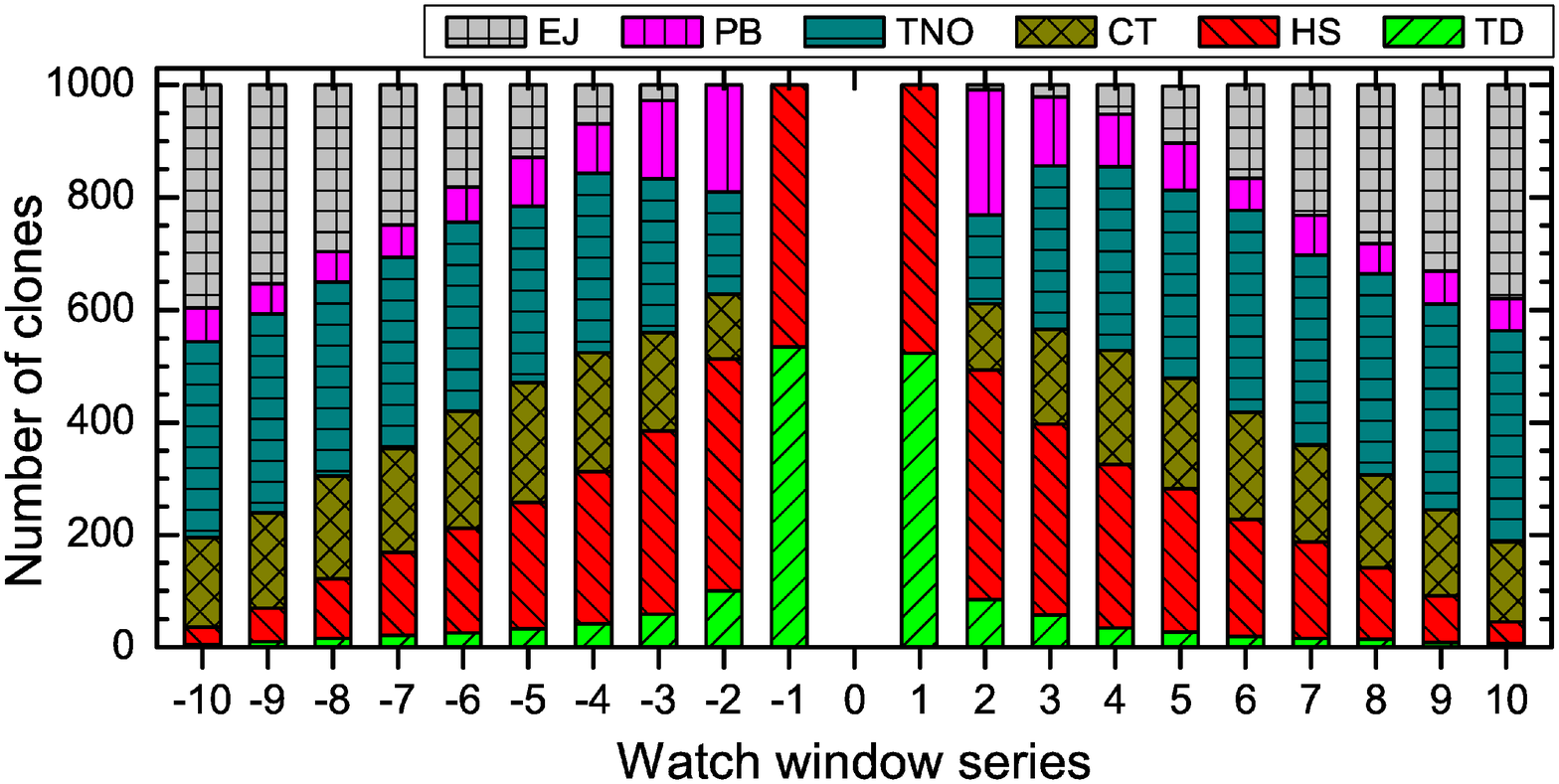}
   \caption{States of the 1000 clones in ten watch
   windows (50\,kyr each) evenly spreading in the 10\,Myr duration of
   both the forward and backward integrations. The series numbers of
   windows are in the temporal sequence. }
   \label{kvcat}
   \end{figure}

Based on the behaviors of two variables, namely the semimajor axis
$a$ and resonant angle $\sigma$, we divide the clone orbits into
seven categories:
\begin{itemize}
  \item Tadpole orbits, TD for abbreviation
  \item Horseshoe orbits, HS for abbreviation
  \item Retrograde satellite, RS for abbreviation
  \item Centaurs, CT for abbreviation
  \item Transneptunian objects\footnote{Roughly speaking, Transneptunian
Objects (also known as the ``Kuiper belt objects'') are those
celestial objects whose semimajor axes are larger than that of
Neptune; and Centaurs are celestial objects with semimajor axes
between that of Jupiter and Neptune.}, TNO for abbreviation
  \item Passing by orbits, PB for abbreviation
  \item Ejected objects, EJ for abbreviation
\end{itemize}
Actually, the first three all belong to co-orbital motions. They
have similar $a$, but differ from each other by the libration center
of $\sigma$: $\pm 60^\circ$ for TD, $\pm 180^\circ$ for HS, and
$0^\circ$ for RS. As for Centaurs and TNOs, our definition is not so
rigid as in common sense. For a clone whose $\sigma$ does not
librate, we simply check if its semimajor axis is larger or smaller
than the upper or lower boundary of Trojan semimajor axis. This
boundary is estimated through $\frac{d}{D}\approx \sqrt{3\mu}\,a_8$
where $d$ and $D$ represent the amplitudes of $a$ and $\sigma$,
respectively, and $a_8$ is the semimajor axis of Neptune. Here the
angular boundary $D$ is set to be $70^{\circ}$ according to our
previous study \citep{zly1}, and the upper and lower boundaries of
$a$ are then defined by $a_8 \pm d$. The orbits located out of the
upper boundary are regarded as TNOs, while CTs are those inside the
lower boundary. If in a watch window the semimajor axis of a clone
excurses both sides of the Trojan boundaries, it is assigned to the
PB category. And finally, those with semimajor axis larger than
100\,AU are regarded as being ejected from the system. We set the
watch window to be 50\,kyr, covering about 5 full librating periods
of a tadpole orbit. Over the 10\,Myr's duration of our simulations,
ten evenly distributed windows are set. We check the 1000 clones in
every window and the statistical results are illustrated in
Fig.~\ref{kvcat}.

The forward and backward integration give more or less the same
results in Fig.~\ref{kvcat}. Note that though we find RSs when
closely examining the orbits, no RS appear in Fig.~\ref{kvcat}.
Generally the RS phase for a clone orbit lasts only for a very short
duration, and it cannot be recognized by our numerical
categorization code. On one hand, we need longer window width to
assure the orbits type; on the other hand, we need shorter time
interval to avoid type mixing in a window. Surely such a dilemma
also causes deviations to other categories, but not too much.

As Fig.~\ref{kvcat} shows, the number of PBs and all coorbital
orbits (TD and HS) decreases with time, indicating that these
orbital types are only temporary phases, and not likely to be the
final destiny of 2004 KV18. While the portion of TNOs and CTs seem
not change much after $3\times 10^5$\,yr, implying that they may be
the potential final destinies for 2004 KV18. Since the number of CTs
is smaller than that of TNOs, and this number decreases slightly
with time, we argue that it's more possible that 2004 KV18 was ever
and will end as a TNO than as a Centaurs.

\section{2008 LC18}
At first glance, the asteroid 2008 LC18 is more like a typical NT
than 2004 KV18, because it has a small eccentricity. Locating the
nominal orbit $(a,e,i)$ at epoch JD=2449200.5 on the dynamical maps
in our previous papers \citep[figure~3 in][]{zly2}, we know that it
is on the edge of stable region. The {\it spectral
number}\footnote{The spectral number is an indicator indicating the
regularity (or, stability) of an orbit. A regular (thus stable)
orbit has a small SN while a chaotic (unstable) orbit has a large
SN. See our previous paper \citep{zly1} for the definition.} (SN) of
it is $\sim 50$. This SN is nearly the same as the ones for the
$L_4$ NTs, 2005 TO74 and 2001 QR322, but is larger than the SNs for
other four $L_4$ NTs (2004 UP10, 2005 TN53, 2006 RJ103 and 2007
VL305). So the stability of 2008 LC18 is comparable to 2005 TO74 and
2001 QR322, whose orbits have been discussed in our previous paper
\citep{zly2}. As a matter of fact, in the paper reporting the
detection of 2008 LC18, \cite{she10a} mentioned that all the orbits
of known NTs by then (seven NTs except for 2004 KV18) ``are stable
for the age of the Solar system''. On the contrary, 2004 KV18 just
discussed in previous section, has an ${\rm SN} > 100$, indicating a
very chaotic thus unstable orbit.

The orbit of 2008 LC18 is less precisely determined compared to 2004
KV18, as indicated by the larger $1\sigma$ variations in
Table~\ref{orbele}. Using the same method for 2004 KV18, a cloud of
1000 clone orbits is generated around the nominal orbit in the
6-dimensional orbital elements space. The initial conditions
($a,e,i$ and $\sigma$) of these clones can be seen in
Fig.~\ref{lcini}. We integrate these clone orbits for $4\times
10^9$\,yr (4\,Gyr, roughly the lifetime of the Solar system) in both
forward and backward directions. The results are presented below.

\subsection{Initial conditions and orbital stability}

We monitor the resonant angle $\sigma$ of each clone orbit in the
numerical simulations. When $\sigma < -180^\circ$ for the first time
($t_1$), the identity of the clone as an $L_5$ NT is regarded as
ended. In our calculations, we found that as soon as a clone
deviates from the tadpole orbit it will leave the 1:1 MMR very soon
later. Therefore, we ignore the moment of escaping from the 1:1 MMR,
i.e. $t_2$, but focus only on $t_1$ in this part.

 \begin{figure}
   \centering
   \includegraphics[width=\textwidth, height=6.5cm, angle=0]{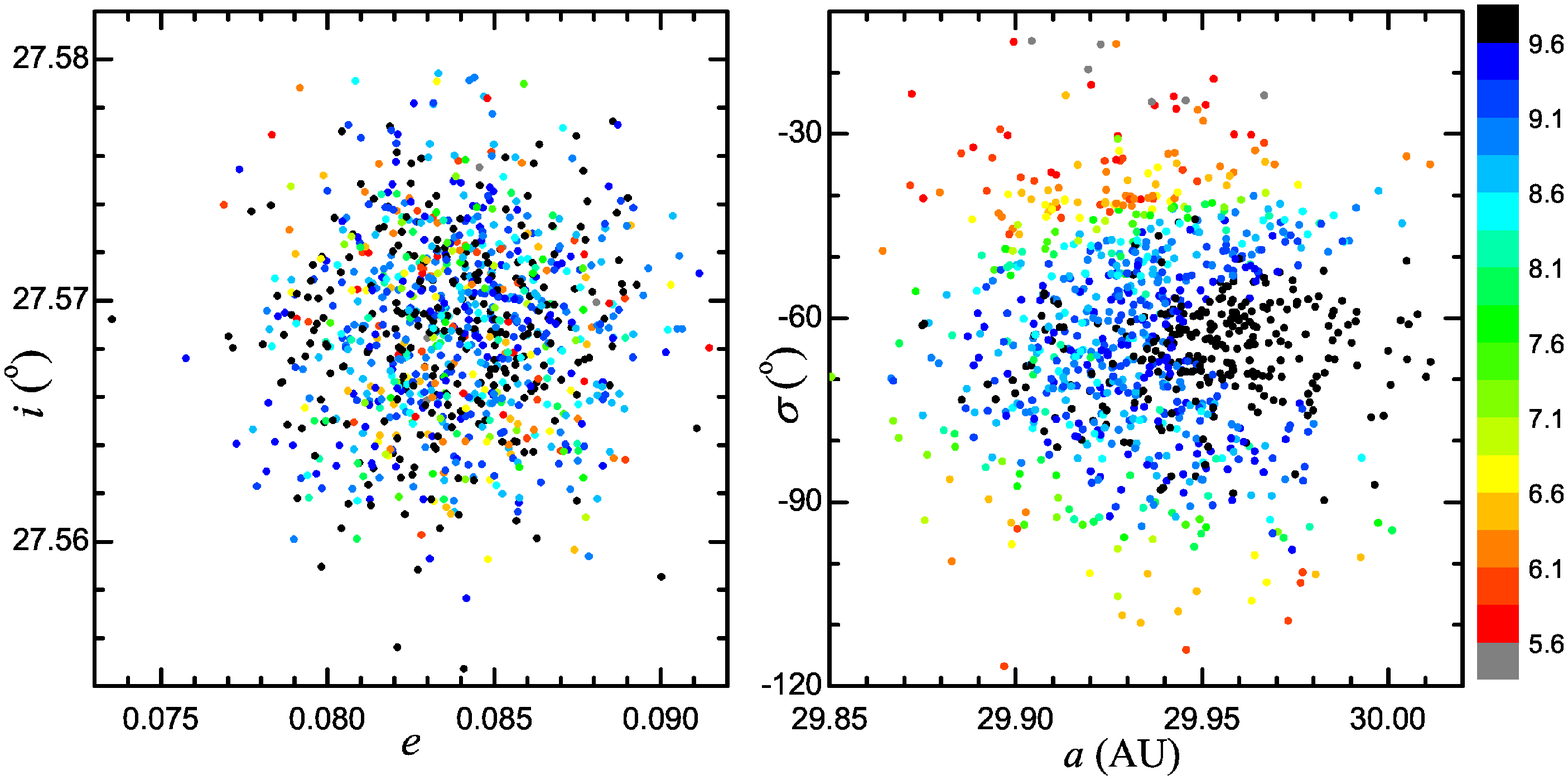}
   \caption{The initial conditions of clone orbits of 2008 LC18. The color
   code indicates the lifetime of the clone as an $L_5$ NT (as the $t_1$
   in Fig.~\ref{kvesct}, in logarithmical scale). Those orbits staying on
   the $L_5$ Trojan orbits for the whole simulation timespan (4\,Gyr) are
   indicated by black dots. }
   \label{lcini}
 \end{figure}

\begin{figure}
   \centering
   \includegraphics[width=\textwidth, height=6.5cm, angle=0]{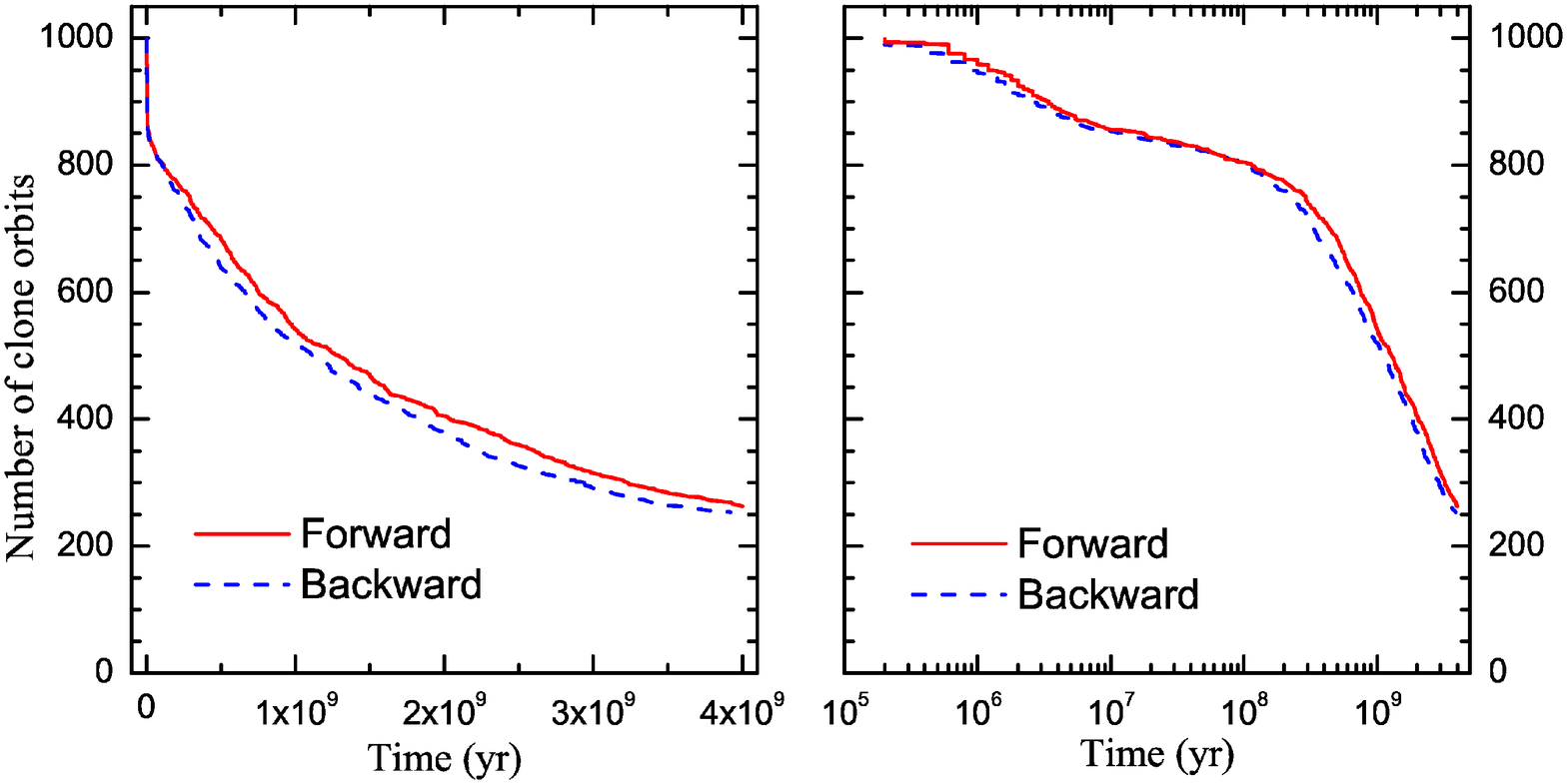}
   \caption{The number of clones that survive in the $L_5$ cloud decreases
   with respect to time. In the left panel, the time is in linear scale
   while in the right panel it's in logarithmical scale. The solid red
   curves are for the forward integrations, and the dashed blue curves for
   the backward integrations.}
   \label{orbnum}
 \end{figure}

Among the 1000 clones, for both forward and backward integrations,
the first escaping from the $L_5$ Trojan region happens at $2\times
10^5$\,yr, but most of clones survive in the $L_5$ cloud beyond
$10^8$\,yr. In Fig.~\ref{lcini} we show the dependence of the
lifetimes ($t_1$) of clone orbits on the initial conditions. Since
our calculations reveal a good temporal symmetry, i.e. the results
from the backward integrations are nearly the same as the results
from the forward integrations (it can be seen clearly in
Fig.~\ref{orbnum} too), we show in Fig.~\ref{lcini} only the case
for the forward integration.

All points with different colors, including the black points
indicating the most stable orbits surviving 4\,Gyr on the Trojan
orbits, spread uniformly in the $(e,i)$ cloud of initial conditions
in the left panel of Fig.~\ref{lcini}. This uniform distribution
indicates that within the orbital elements' uncertainty ranges, the
lifetime ($t_1$) has no relation to either the initial eccentricity
$e$ or the initial inclination $i$. We also check carefully the
dependence of lifetime on the initial angular variables ($\omega,
\Omega$, $M$) and find that the stability does not have apparent
relevance to these angular orbital elements either (within the error
ranges, of course). Nevertheless, the stability depends on the the
summation of $\omega, \Omega$ and $M$, namely the mean longitude
$\lambda = \omega + \Omega + M$. More precisely, it depends on the
resonant angle $\sigma = \lambda -\lambda_8$, as shown in the right
panel of Fig.~\ref{lcini}.

Also in this picture, the dependence on the semimajor axis is
manifest. The color points make a layered structure. The most
unstable orbits occupy the outer layer to the left, especially, we
find that both too large $(\sigma > -40^\circ)$ and too small
$(\sigma <-100^\circ)$ initial $\sigma$ lead to unstable motion.
Meanwhile, nearly all the most stable orbits concentrate in the
inner layer to the right. The gathered black points form a triangle
in the region of middle $\sigma (\sim -65^\circ)$ and large $a$
($>29.93$\,AU), and no color points exist in this triangle.

In fact, we know from our previous studies \citep{zly1} that the
libration amplitude of $\sigma$ is related to the initial semimajor
axis. The further the initial $a$ is away from the resonant center,
the larger the $\sigma$ amplitude is. Thus, the layered structure in
the right panel may be equivalent to the vertical stripe structure
in the dynamical maps on the $(a,e)$ plane \citep[figure 3
in][]{zly2}, where the initial $\sigma$ is fixed. The ``C type''
secular resonances \citep{zly2} were found to be responsible for the
vertical structures. Recalling the position of 2008 LC18 on the
$(a,e)$ and $(a,i)$ planes, we find that layered structure on the
$(a,\sigma)$ plane in Fig.~\ref{lcini} is probably due to one of the
``C type'' resonance: $4f_\sigma -2 f_{2:1}+g_6+g_7= 0$.

The probability of finding Neptune Trojans in the stable region
should be much higher than in the unstable region, thus the 2008
LC18 is expected to be on a stable orbit. We would argue that
further observations in future will constrain its orbit to the
stable region, particularly, the semimajor axis and the
corresponding resonant angle to the triangle on the $(a,\sigma)$
plane in Fig.~\ref{lcini}, as we mentioned above.

Starting from the 1000 clones and after 4\,Gyr's orbital evolution,
there are still 262 clones surviving in the $L_5$ region ($t_1 >
4$\,Gyr) in the forward integration. As for the backward
integration, 252 orbits survive. In a word, more than 25\% of clones
stay on the tadpole orbit around the $L_5$ point till 4\,Gyr in both
forward and backward directions. The escaping of clones from the
$L_5$ region happens in a wide time range beginning from $2\times
10^5$\,yr. The number of clones that survive on the tadpole orbits
decreases with time. We plot in Fig.~\ref{orbnum} such declines of
clone numbers for integrations of both directions.

Two curves for the forward and backward integrations in
Fig.~\ref{orbnum} coincide with each other, implying the nearly
exact symmetry between two temporal directions. From the profiles of
these curves, two stages of escapes intersecting with each other at
$\sim 10^7$\,yr can be readily recognized. In the first stage, about
$15\%$ of clones ($\sim 150$ clones) escape from the $L_5$ region,
quickly. While in the second stage, the number of surviving clones
decreases slowly. As indicated by the escape times (color) in
Fig.~\ref{lcini}, nearly all the clones that escape in the first
stage have initial resonant angles either $\sigma_0 > -40^\circ$ or
$\sigma_0 < -100^\circ$. The clones escaping in the second stage on
the other hand have $-40^\circ < \sigma_0 < -100^\circ$. In fact, as
the dynamical maps in our previous papers \citep[figure 3
in][]{zly2} show, the stable region is separated from the unstable
region by a sharp edge, i.e. the intermediate area is very narrow.
Thus it's natural to see that those orbits initially locating in the
unstable region escape quickly making the quick decrease in the
first stage in Fig.~\ref{orbnum}, while those orbits in the stable
region escape slowly in the second stage. We believe that 2008 LC18
is a typical NT rather than a temporary NT like 2004 KV18. Further
observations will reduce the uncertainties of the orbit and probably
exclude those unstable clone orbits escaped in the first stage. Then
the decay in the second stage will give the ``proper'' surviving
probability of this objects in the $L_5$ region.

Moreover, for those orbits initially in the stable region, their
orbital elements may diffuse very slowly on the dynamical map and
the instability can set in ``abruptly'' when the orbit crosses the
narrow transitional area. Such an example of sudden escape from the
$L_5$ region is shown in the following section.

\subsection{Orbital evolution}

\begin{figure}
  \centering
   \includegraphics[width=\textwidth, angle=0]{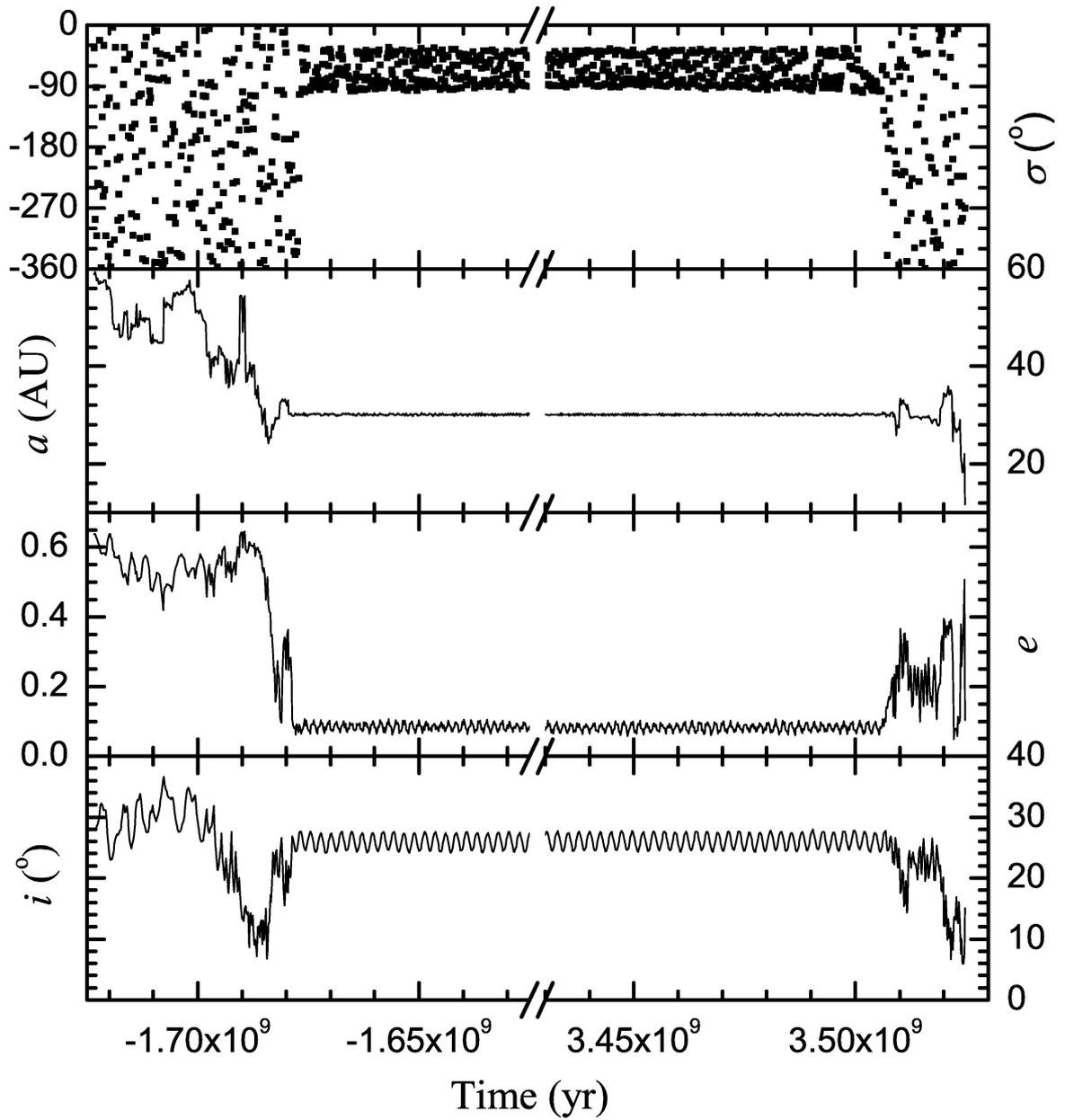}
   \caption{Temporal evolution of the resonant angle ($\sigma$),
   semimajor axis ($a$), eccentricity ($e$) and inclination ($i$)
   of the nominal orbit of 2008 LC18. Note the break in
   abscissa axis.
   }
  \label{lcorb}
\end{figure}

We have shown in the above section the ensemble behavior of clone
orbits, and we will turn to the evolution of individual orbits now.
But in fact, the orbital evolution of clones of 2008 LC18 is plain.
As an example, we illustrate in Fig.~\ref{lcorb} the temporal
evolution of the nominal orbit of 2008 LC18.

From the behavior of resonant angle $\sigma$ in the top panel of
Fig.~\ref{lcorb}, the nominal orbit will leave the $L_5$ Trojan
region in about $3.505$\,Gyr, and it was captured to the $L_5$
tadpole orbit about $1.678$\,Gyr ago. During its being an $L_5$
Trojan, the semimajor axis, eccentricity and inclination behave very
regularly, varying with small amplitudes. Particularly, the
amplitude of $\sigma$ is smaller than $40^\circ$ in most time of its
life. No apparent secular variations of $a, e, i$ or $\sigma$ can be
observed in the time range when it is on the tadpole orbit as a
Trojan. As we mentioned above, the chaos seems to set in suddenly.
This is due to the fact that the border between the stable and
unstable region is narrow, a small deviation from the stable region
may result in destroying of the stability.

Actually, all those stable orbits that survive on the tadpole orbits
in both time directions for 4\,Gyr behave in a similarly regular way
as the nominal orbit in the Trojan phase from $-1.678$\,Gyr to
$3.505$\,Gyr. Their semimajor axes oscillate between 29.98\,AU and
30.4\,AU, their eccentricities are always smaller than 0.12, their
inclinations librate around $26^\circ$ with amplitudes smaller than
$4^\circ$, and their resonant angles librate around $-60^\circ$ with
amplitudes $\sim 30^\circ$.

The second panel of Fig.~\ref{lcorb} shows that the semimajor axis
of the nominal orbit was beyond the orbit of Neptune before it was
captured into the Trojan cloud, and it will be inside Neptune's
orbit after it leaves the $L_5$ region. Therefore, it had become a
NT from a scattered Transneptunian Objects (TNO) and it will turn to
a Centaurs finally in future. However, it is worth noting that we
may draw conclusions from our calculations about the destiny of
these clones 4\,Gyr later in future, but no solid conclusions can be
made about their origins 4\,Gyr ago in the past. The Solar system
currently is quite ``clean'', but 4\,Gyr ago the orbits of planets
probably had not been settled down and the planetesimal disc or even
the gas disc still existed. Even though the computing error
(roundoff error and model error) could be ideally controlled and
ignored absolutely, the origin of a NT cannot be determined by
tracing back its orbit through backward integration, because the
exact circumstances at that time are far from being understood.

\subsection{Inclination excited {\it in situ}?}

Although the eccentricity of 2008 LC18 is small ($\sim 0.08$), its
inclination is quite high ($\sim 27.5^\circ$), suggesting a typical
excited orbit. The high inclination of some NTs is an important
puzzle and perhaps also a key clue for the understanding of their
origins and orbital evolutions. Based on our calculations for those
1000 clones, we discuss in this part the excitation of NT orbits,
particularly, we will check the possibility of inclination being
pumped up {\it in situ}.

\begin{figure}
   \centering
   \includegraphics[width=\textwidth, height=6.5cm, angle=0]{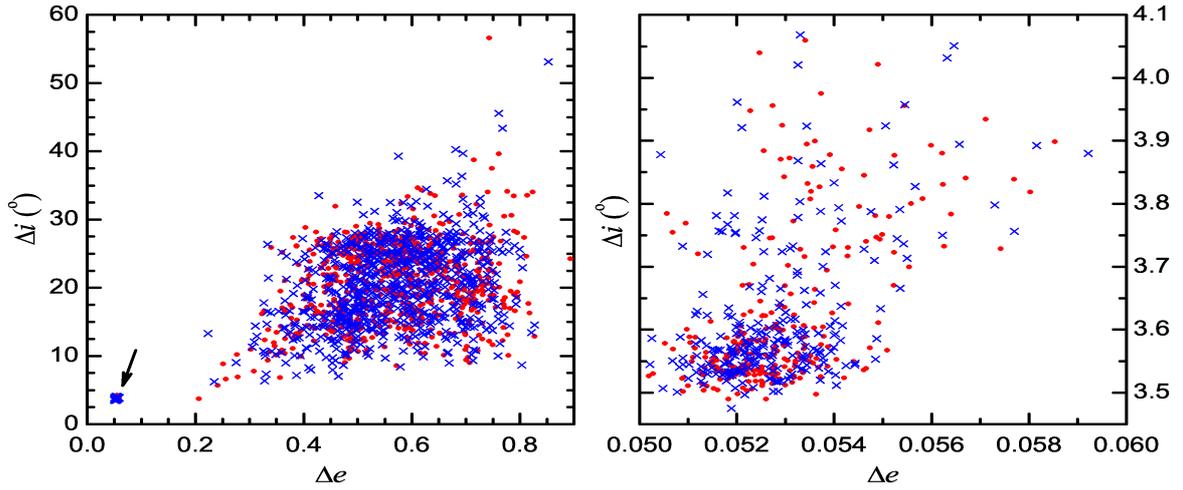}
   \caption{The variations of the eccentricity and inclination of clone orbits.
   The red solid circles are for the forward integrations and the blue crosses
   for backward integrations. The right panel shows the variations for those
   stable orbits surviving in the $L_5$ region for 4\,Gyr. It is an enlargement
   of the lower left corner of the left panel, indicated by an arrow (see text). }
   \label{varei}
 \end{figure}

The resonant maps \citep[figures 5, 6, 7 in][]{zly2} reveal that
numerous secular resonances involve in the orbital dynamics of NTs,
even in the stable region of the orbital elements space. The effects
of these resonances may be not so strong that the orbital stability
of NTs may not be destroyed in the age of the Solar system. But they
may drive the orbits of NTs to diffuse slowly in the elements space,
so that a primordial NT may attain the high inclination through this
slow diffusion. To clarify wether this process is responsible for
the inclination of 2008 LC18, we examine the eccentricity variation
$\Delta e$ and inclination variation $\Delta i$ of clone orbits
during the simulations. Here $\Delta e$ and $\Delta i$ are the full
varying ranges of eccentricity and inclination, defined as the
differences between the maximal and minimal values of $e$ and $i$ in
the whole simulations. The results are summarized in
Fig.~\ref{varei}.

In the left panel of Fig.~\ref{varei}, clone orbits are divided into
two separated groups. One group extends from $\Delta e=0.2$ to
$\Delta e=0.85$, and the other one is confined in a small area at
low $\Delta e$ and low $\Delta i$, indicated by the arrow in the
picture. We pick out all the stable orbits surviving 4\,Gyr in our
simulations in both integration directions and plot their variations
of $e$ and $i$ in the right panel of Fig.~\ref{varei}. Apparently,
the above mentioned group of orbits locating in the lower left
corner of the left panel just consists of those stable orbits
exactly. So, all the stable orbits have small eccentricity and
inclination variations, i.e. $0.05 < \Delta e < 0.06$ and
$3.45^\circ < \Delta i < 4.10^\circ$. For those stable orbits, the
inclination neither decreases significantly in the backward
integration nor increases in the forward integration. Hence the
inclination of 2008 LC18 seemingly has not arisen under the current
configuration of the Solar system.

However, those unstable orbits occupy the lower right half of the
left panel in Fig.~\ref{varei}, reflecting that most of them
experience large eccentricity variations. A carefully examining on
the orbits reveals that the eccentricity variation is caused either
by crossing through MMRs (between clones and planets) or by close
encounters of the clones with planets. And both of these two
mechanisms work only when the clones are outside the Trojan phase.
Generally, the increased eccentricity of a clone will result in
further close encounters with planets. The inclination may obtain
large variation in these encounters. That is the reason why large
inclination variation is always accompanied by large eccentricity
variation. On the contrary, the lack of orbits in the upper left
region of the picture shows clearly that no inclination excitation
could happen without the eccentricity excitation.

To the question of whether the inclination of 2008 LC18 can be
excited {\it in situ}, our calculations give a negative answer. But
our calculations suggest that during the capturing of asteroids into
the Trojan orbits, their inclinations might be pumped up, mainly
through close encounters with planets. After the capturing, there
must be some braking mechanisms that can damp down the eccentricity
but preserve the inclination.

\section{Conclusions}

\label{sect:conclusion}

The Trojan cloud around the triangular Lagrange points $L_4, L_5$ of
Neptune is believed to be a large reservoir of asteroids, hosting
more asteroids than Jupiter's Trojan cloud or even the main asteroid
belt between Mars and Jupiter \citep{she06,she10b}. Up to now, six
leading Neptune Trojans around the $L_4$ point and two trailing ones
around the $L_5$ point have been discovered. The dynamics of the
$L_4$ NTs have been studied in literatures by several authors
\cite[e.g.][]{mar03,bra04,lij07,hor10}. In this paper, taking into
account the errors introduced in the observations and orbital
determinations, we investigate the orbital dynamics of two trailing
NTs, 2004 KV18 and 2008 LC18. Starting from clouds of clone orbits
around the nominal orbits, we simulate the clones' orbital
evolutions using the well-known {\it Mercury6} numerical integrator
package.

Our results suggest that 2004 KV18 is on an especially unstable
orbit. It is neither a primordial nor a permanent NT, but rather a
passer-by object on its way of exchanging between a TNO and a
Centaurs. Its lifetime as a trailing NT is in the order of
$10^5$\,years in the future, and probably it has been on such a
unstable tadpole orbit only for less than $2\times 10^5$\,years.
Such an unstable orbit means that it can neither be regard as the
smoking gun of ``the hot Trojan'' from the chaotic capture model
\citep{nes09} nor from the migrating Neptune model
\citep{lyk09,lyk10a,lyk10b}.

Due to the strong chaos suffered by the orbits, it is hard to draw
solid conclusion about where the asteroid came from and where it
will go in the long term. But statistics on the clone orbits still
give some helpful information, suggest that most probably it will
evolve to be a TNO after its leaving the $L_5$ region.


The asteroid 2008 LC18 however is more like a primordial trailing
Neptune Trojan. Our calculations show that an appreciable proportion
of clone orbits within the limits of orbital errors survives on the
tadpole orbits for 4\,Gyr and their orbital evolutions are very
regular in both forward and backward time directions. Particularly,
the high inclination of the orbit does not change much, implying
that the orbit has not been excited on the tadpole orbit under
current planetary configuration. On the contrary, for those clone
orbits escaping from the tadpole orbit, their inclinations may vary
significantly when they are outside of the Trojan phase. These
calculations imply that the 2008 LC18 may be captured onto current
high-inclined orbit very long ago, and during the capturing process
its inclination may be excited due to close encounters with planets.

The orbital stability of clones of 2008 LC18 apparently depends on
the semimajor axis and the resonant angle. Nearly all the stable
orbits are in a specific region of the $(a,\sigma)$ plane, so we
expect that additional observations will confine its orbit into this
region.

\section*{Acknowledgements} This work was supported by the Natural
Science Foundation of China (NSFC, No.~10833001, No.~11073012) and
by Qing Lan Project (Jiangsu Province). J.~Li is also supported by
NSFC under grants No.~1103008 and No.~11078001.

\label{lastpage}


\begin{thebibliography}{99}
\small \setlength{\itemindent}{-3mm} \setlength{\itemsep}{-0.5mm}
\setlength{\baselineskip}{4.5mm}



\bibitem[Bowell et al. (1990)]{bow90}
Bowell E., Holt H. E., Levy D. H., Innanen K. A., Mikkola S.,
Shoemaker E. M., 1990, BAAS, 22, 1357

\bibitem[Braser et al. (2004)]{bra04}
Braser R., Mikkola S., Huang T.-Y., Wieggert P., Innanen K., 2004,
MNRAS, 347, 833

\bibitem[Chambers (1999)]{cham99}
Chambers J., 1999, MNRAS, 304, 793

\bibitem[Chiang et al. (2003)]{chi03}
Chiang E. I., Jordan A. B., Millis R. L., et al., 2003, AJ, 126, 430

\bibitem[Connors et al. (2011)]{con11}
Connors M., Wieggert P., Veillet C. 2011, Nature, 475, 481

\bibitem[Crida (2009)]{cri09}
Crida A., 2009, arXiv:0903.3008

\bibitem[Duncan et al. (1989)]{dun89}
Duncan M., Quinn T., Tremaine S., 1989, Icarus, 82, 402

\bibitem[Dvorak et al. (2007)]{dvo07}
Dvorak R., Schwarz R. S\"uli \'A, Kotoulas T., 2007, MNRAS, 382,
1324

\bibitem[Dvorak et al. (2010)]{dvo10}
Dvorak R., Bazso A., Zhou L.-Y., 2010, Celest. Mech. Dyn. Astron.,
107, 51

\bibitem[Dvorak et al. (2012)]{dlz12}
Dvorak R., Lhotka C., Zhou L.-Y., 2012, A\&A, DOI:
10.1051/0004-6361/201118374

\bibitem[Horner \& Lykawka (2010)]{hor10}
Horner J.,Lykawka P.S.,2010, MNRAS, 405,49

\bibitem[Li et al. (2007)]{lij07}
Li J., Zhou L.-Y., Sun Y.-S., 2007, A\&A, 464, 775

\bibitem[Lykawka et al. (2009)]{lyk09}
Lykawka P., Horner J., Jones B., Mukai T., 2009, MNRAS, 398, 1715

\bibitem[Lykawka et al. (2010)]{lyk10a}
Lykawka P., Horner J., Jones B., Mukai T., 2010, MNRAS, 404, 1272


\bibitem[Lykawka \& Horner (2010)]{lyk10b}
Lykawka P.S., Horner J., 2010, MNRAS, 405, 1375

\bibitem[Mainzer et al. (2011)]{mai11}
Mainzer, A., Bauer, J., Grav, T. and 32 coauthors, 2011, ApJ, 731,
53

\bibitem[Marzari et al. (2003)]{mar03}
Marzari F., Tricarico P., Scholl H., 2003, A\&A, 410, 725

\bibitem[Michtchenko \& Ferraz-Mello (1995)]{mit95}
Michtchenko T., Ferraz-Mello S., 1995, A\&A, 303, 945

\bibitem[Morbidelli et al. (2005)]{mor05}
Morbidelli A., Levison H.F., Tsiganis K., Gomes R., 2005, Nature,
435 462

\bibitem[Murray \& Dermott (1999)]{mur99}
Murray C.D., Demott S. F., 1999, Solar System Dynamics(New York:
Cambridge Univ. Press)

\bibitem[Nesvorn\'y \& Dones (2002)]{nes02a}
Nesvorn\'y D., Dones L., 2002, Icarus, 160, 271

\bibitem[Nesvorn\'y \& Vokrouhlick\'y (2009)]{nes09}
Nesvorn\'y D., Vokrouhlick\'y D., 2009, AJ, 137, 5003

\bibitem[Robutel \& Gabern (2006)]{rob06}
Robutel P., Gabern F., 2006, MNRAS, 372, 1463

\bibitem[Robutel \& Bodossian (2009)]{rob09}
Robutel P., Bodossian J., 2009, MNRAS, 399, 69


\bibitem[Sheppard \& Trujillo (2006)]{she06}
Sheppard S., Trujillo C., 2006, Science, 313, 511

\bibitem[Sheppard \& Trujillo (2010a)]{she10a}
Sheppard S., Trujillo C., 2010, Science, 329, 1304

\bibitem[Sheppard \& Trujillo (2010b)]{she10b}
Sheppard S., Trujillo C., 2010, ApJ, 723, L233


\bibitem[Wisdom (1980)]{wis80}
Wisdom J., 1980, AJ, 85, 1122

\bibitem[Zhou et al. (2009)]{zly1}
Zhou L.-Y., Dvorak R., Sun Y.-S., 2009, MNRAS, 398, 1217

\bibitem[Zhou et al. (2011)]{zly2}
Zhou L.-Y., Dvorak R., Sun Y.-S., 2011, MNRAS, 410, 1849


\end{thebibliography}
\end{document}